\documentclass{aa}

\usepackage{graphicx}
\usepackage{upgreek}
\usepackage{color}

\usepackage[varg]{txfonts}
\usepackage{microtype}
\usepackage{booktabs}
\usepackage{xcolor}
\usepackage{hyperref}
\usepackage{graphicx}
\usepackage{multirow}
\usepackage{float}
\usepackage{adjustbox}

\usepackage{hhline}
\hypersetup{%
  colorlinks=true, % hyperlinks will be black
  linkcolor=blue, % hyperlink borders will be red
  citecolor=blue
}

\usepackage{etoolbox}
\makeatletter
\let\oldcitet=\citet

\renewcommand{\citet}[1]{\textcolor[rgb]{0,0,1}{\oldcitet{#1}}}

\newcommand{\tablenotea}[1]{\parbox{15.4cm}{\indent \footnotesize{#1}}}

\begin{document}

\title{Quantum study of reaction O\,($^3P$)\,+\,H$_2$\,($v,j$)\,$\rightarrow$\,OH\,+\,H:\\
OH formation in strongly UV-irradiated gas}

%\subtitle{OH formating in strongly irradiated PDRs}

\titlerunning{Quantum study of reaction O\,($^3P$)\,+\,H$_2$\,($v,j$)}
\authorrunning{Veselinova et al.}

\author{
A.~Veselinova\inst{1,2},  M.~Ag\'undez\inst{3},  J.~R.~Goicoechea\inst{3}, M.~Men\'endez\inst{2}, A.~Zanchet\inst{3}, E.~Verdasco\inst{2}, P.~G.~Jambrina\inst{1}, and F.~J.~Aoiz\inst{2}
}

\institute{
$^1$ Departamento de Qu\'imica F\'isica, University of Salamanca, Plaza Caidos S/N, E-37008, Salamanca, Spain.\\
\email{pjambrina@usal.es}\\
$^2$ Departamento de Qu\'imica F\'isica (Unidad Asociada al CSIC), Universidad Complutense de Madrid, Ciudad Universitaria, S/N, E-20840, Madrid, Spain.\\
\email{aoiz@quim.ucm.es}\\
$^3$ Instituto de F\'{\i}sica Fundamental, CSIC, Calle Serrano 121-123, E-28006, Madrid, Spain.\\
\email{marcelino.agundez@csic.es, alexandre.zanchet@csic.es}\\
}

\date{20 January  2021 / Accepted xx, xx, 2021}

% \abstract{}{}{}{}{}

\abstract
{The reaction between atomic oxygen and molecular hydrogen is an important one in astrochemistry as it regulates the abundance of the hydroxyl radical and serves to open the chemistry of oxygen in diverse astronomical environments. However, the existence of a high activation barrier in the reaction with ground state oxygen atoms limits its efficiency in cold gas. In this study we calculate the dependence of the reaction rate coefficient on the rotational and vibrational state of H$_2$ and evaluate the impact on the abundance of OH in interstellar regions strongly irradiated by far-UV photons, where H$_2$ can be efficiently pumped to excited vibrational states. We use a recently calculated potential energy surface and carry out time-independent quantum mechanical scattering calculations to compute rate coefficients for the reaction O\,($^3P$)\,+\,H$_{2}$\,($v,j$)\,$\rightarrow$\,OH\,+\,H, with H$_2$ in vibrational states \mbox{$v$ = 0-7} and rotational states \mbox{$j$ = 0-10}. We find that the reaction becomes significantly faster with increasing vibrational quantum number of H$_2$, although even for high vibrational states of H$_2$ ($v$ = 4-5) for which the reaction is barrierless, the rate coefficient does not strictly attain the collision limit and still maintains a positive dependence with temperature. We implemented the calculated state-specific rate coefficients in the Meudon PDR code to model the Orion Bar PDR and evaluate the impact on the abundance of the OH radical. We find the fractional abundance of OH is enhanced by up to one order of magnitude in regions of the cloud corresponding to $A_{\rm V}$ = 1.3-2.3, compared to the use of a thermal rate coefficient for O + H$_2$, although the impact on the column density of OH is modest, of about 60\%. The calculated rate coefficients will be useful to model and interpret JWST observations of OH in strongly UV-illuminated environments.}


\maketitle

\section{Introduction}

Hydroxyl (OH) is a key radical in interstellar oxygen chemistry. OH is expected to be particularly abundant in \mbox{high-temperature} environments driven by the presence of stellar ultraviolet (UV) \mbox{photons}, so called photodissociation regions \citep[PDRs; e.g.,][]{Sternberg95,Hollenbach09} and in shocked gas \citep[i.e.,][]{Elitzur_1978,Hollenbach79,Draine83,Kaufman96}. In these environments the gas attains the high temperatures (\,$>$\,300\,K) needed to make the endothermic neutral-neutral reaction
\begin{equation} \label{reac-1}
{\rm O\,(^3{\it{P}}) + H_2\,(^1\Sigma^+) \rightarrow OH\,(^2\Pi) + H\,(^2{\it{S}})}
\end{equation}
an important source of OH. Reaction~(\ref{reac-1}) is thought to drive the formation of the OH observed in diffuse clouds \citep[e.g.,][]{Neufeld_2002,Godard14}, protostellar outflows \citep[e.g.,][]{Melnick90,Wampfler13,Goico15}, \mbox{massive winds} from extragalactic nuclei \citep[e.g.,][]{Sturm11,Gonzalez12},  the \mbox{UV-illuminated}  rims of molecular clouds \citep[][]{Goicoechea11,Parikka17}, and the upper layers of protoplanetary disks \citep{Mandell08,Fedele13}. In addition, \mbox{H$_2$O photodissociation}  contributes to  the production of OH \citep[see][]{Tabone21} in environments that host  high water vapor abundances and intense \mbox{far-UV} (FUV; photon energy $<$\,13.6\,eV) radiation fluxes \mbox{\citep[e.g.,][]{Bergin03,Tappe08,Karska13}}.

Many hydrogen abstraction reactions \mbox{X\,+\,H$_2$\,$\rightarrow$\,XH\,+\,X} \mbox{(e.g., with X\,$=$\,O, C$^+$, C, S$^+$, and S)} are very endothermic and/or have sizable energy barriers \citep[see, e.g.,][]{Gerin16}. Hence, the formation of the radical XH through this \mbox{gas-phase} route is expected to be very slow in cold gas interstellar conditions. However, strongly illuminated PDRs such as the Orion Bar \citep[an edge-on rim of the Orion molecular cloud; e.g.,][]{Tielens93,Goicoechea16} are characterized by both warm gas temperatures and enhanced abundances of \mbox{FUV-pumped} vibrationally-excited molecular hydrogen \citep[up to \mbox{$v$\,=\,10}; see, e.g.,][]{Kaplan17}. It was long suspected that the presence of large column densities of \mbox{FUV-pumped} H$_2$\,($v$\,$\geq$1) could increase the reactivity of some of the above hydrogen abstraction reactions \citep{Stecher72,Freeman82,Tielens_1985a,Sternberg95}.

\begin{table*}
\caption{Rate coefficient parameters for O($^3P$) + H$_2$\,($v,j$).} \label{table:rates}
\small
\centering
\begin{tabular}{ccccccccccc}
\hline \hline
\multicolumn{2}{c}{Initial H$_2$ state} & \multicolumn{1}{c}{$\alpha$ (cm$^3$ s$^{-1}$)}  & \multicolumn{1}{c}{$\beta$} & \multicolumn{1}{c}{$\gamma$ (K)} & \hspace{1cm} & \multicolumn{2}{c}{Initial H$_2$ state} & \multicolumn{1}{c}{$\alpha$ (cm$^3$ s$^{-1}$)}  & \multicolumn{1}{c}{$\beta$} & \multicolumn{1}{c}{$\gamma$ (K)}  \\
\hline
\multirow{11}{*}{$v=0$} & \multicolumn{1}{c}{ $j=0$} & \multicolumn{1}{c}{    $4.873 \times 10^{-14}$  } & \multicolumn{1}{c}{  2.9453  } & \multicolumn{1}{r}{  3145.4 } & & \multirow{11}{*}{$v=4$}   & \multicolumn{1}{c}{ $j=0$} & \multicolumn{1}{c}{    $1.269 \times 10^{-11}$  } & \multicolumn{1}{c}{  1.2721  } & \multicolumn{1}{r}{  87.9   }\\\cline{2-5} \cline{8-11}
& \multicolumn{1}{c}{ $j=1$} & \multicolumn{1}{c}{    $3.849 \times 10^{-14}$  } & \multicolumn{1}{c}{  3.0578  } & \multicolumn{1}{r}{  2827.4 }   & & & \multicolumn{1}{c}{ $j=1$} & \multicolumn{1}{c}{    $2.537 \times 10^{-11}$  } & \multicolumn{1}{c}{  0.9839  } & \multicolumn{1}{r}{  181.8  }\\\cline{2-5} \cline{8-11}
& \multicolumn{1}{c}{ $j=2$} & \multicolumn{1}{c}{    $2.975 \times 10^{-14}$  } & \multicolumn{1}{c}{  3.2066  } & \multicolumn{1}{r}{  2378.9 }   & & & \multicolumn{1}{c}{ $j=2$} & \multicolumn{1}{c}{    $2.834 \times 10^{-11}$  } & \multicolumn{1}{c}{  0.9737  } & \multicolumn{1}{r}{  116.5  }\\\cline{2-5} \cline{8-11}
& \multicolumn{1}{c}{ $j=3$} & \multicolumn{1}{c}{    $1.642 \times 10^{-14}$  } & \multicolumn{1}{c}{  3.5176  } & \multicolumn{1}{r}{  1818.9 }   & & & \multicolumn{1}{c}{ $j=3$} & \multicolumn{1}{c}{    $2.708 \times 10^{-11}$  } & \multicolumn{1}{c}{  1.0343  } & \multicolumn{1}{r}{  75.0   }\\\cline{2-5} \cline{8-11}
& \multicolumn{1}{c}{ $j=4$} & \multicolumn{1}{c}{    $6.650 \times 10^{-15}$  } & \multicolumn{1}{c}{  3.9662  } & \multicolumn{1}{r}{  1224.9 }   & & & \multicolumn{1}{c}{ $j=4$} & \multicolumn{1}{c}{    $2.721 \times 10^{-11}$  } & \multicolumn{1}{c}{  1.0737  } & \multicolumn{1}{r}{  123.5  }\\\cline{2-5} \cline{8-11}
& \multicolumn{1}{c}{ $j=5$} & \multicolumn{1}{c}{    $7.522 \times 10^{-15}$  } & \multicolumn{1}{c}{  3.9510  } & \multicolumn{1}{r}{  1045.5 }   & & & \multicolumn{1}{c}{ $j=5$} & \multicolumn{1}{c}{    $3.099 \times 10^{-11}$  } & \multicolumn{1}{c}{  1.0361  } & \multicolumn{1}{r}{  156.3  }\\\cline{2-5} \cline{8-11}
& \multicolumn{1}{c}{ $j=6$} & \multicolumn{1}{c}{    $8.392 \times 10^{-15}$  } & \multicolumn{1}{c}{  3.9388  } & \multicolumn{1}{r}{  912.4  } & & & \multicolumn{1}{c}{ $j=6$} & \multicolumn{1}{c}{    $3.686 \times 10^{-11}$  } & \multicolumn{1}{c}{  0.9562  } & \multicolumn{1}{r}{  111.5  }\\\cline{2-5} \cline{8-11}
& \multicolumn{1}{c}{ $j=7$} & \multicolumn{1}{c}{    $1.097 \times 10^{-14}$  } & \multicolumn{1}{c}{  3.8553  } & \multicolumn{1}{r}{  848.8  } & & & \multicolumn{1}{c}{ $j=7$} & \multicolumn{1}{c}{    $4.902 \times 10^{-11}$  } & \multicolumn{1}{c}{  0.8408  } & \multicolumn{1}{r}{  103.4  }\\\cline{2-5} \cline{8-11}
& \multicolumn{1}{c}{ $j=8$} & \multicolumn{1}{c}{    $1.538 \times 10^{-14}$  } & \multicolumn{1}{c}{  3.7356  } & \multicolumn{1}{r}{  793.8  } & & & \multicolumn{1}{c}{ $j=8$} & \multicolumn{1}{c}{    $6.081 \times 10^{-11}$  } & \multicolumn{1}{c}{  0.7665  } & \multicolumn{1}{r}{  95.3   }\\\cline{2-5} \cline{8-11}
& \multicolumn{1}{c}{ $j=9$} & \multicolumn{1}{c}{    $1.968 \times 10^{-14}$  } & \multicolumn{1}{c}{  3.6597  } & \multicolumn{1}{r}{  703.5  } & & & \multicolumn{1}{c}{ $j=9$} & \multicolumn{1}{c}{    $7.297 \times 10^{-11}$  } & \multicolumn{1}{c}{  0.7072  } & \multicolumn{1}{r}{  84.2   }\\\cline{2-5} \cline{8-11}
& \multicolumn{1}{c}{ $j=10$} & \multicolumn{1}{c}{    $2.823 \times 10^{-14}$ } & \multicolumn{1}{c}{   3.5505 } & \multicolumn{1}{r}{   662.5 } & & & \multicolumn{1}{c}{ $j=10$} & \multicolumn{1}{c}{    $9.703 \times 10^{-11}$ } & \multicolumn{1}{c}{   0.6053 } & \multicolumn{1}{r}{   148.6 }\\\cline{2-5} \cline{8-11}
\hline
\multirow{11}{*}{$v=1$} & \multicolumn{1}{c}{ $j=0$} & \multicolumn{1}{c}{    $5.833 \times 10^{-15}$  } & \multicolumn{1}{c}{  4.1903  } & \multicolumn{1}{r}{  354.2  } & & \multirow{11}{*}{$v=5$} & \multicolumn{1}{c}{ $j=0$} & \multicolumn{1}{c}{    $3.758 \times 10^{-11}$  } & \multicolumn{1}{c}{  0.8816  } & \multicolumn{1}{r}{  -63.7  }\\\cline{2-5} \cline{8-11}
& \multicolumn{1}{c}{ $j=1$} & \multicolumn{1}{c}{    $1.504 \times 10^{-14}$  } & \multicolumn{1}{c}{  3.7943  } & \multicolumn{1}{r}{  484.4  } & & & \multicolumn{1}{c}{ $j=1$} & \multicolumn{1}{c}{    $4.797 \times 10^{-11}$  } & \multicolumn{1}{c}{  0.8114  } & \multicolumn{1}{r}{  $-$7.2   }\\\cline{2-5} \cline{8-11}
& \multicolumn{1}{c}{ $j=2$} & \multicolumn{1}{c}{    $2.956 \times 10^{-14}$  } & \multicolumn{1}{c}{  3.5601  } & \multicolumn{1}{r}{  466.7  } & & & \multicolumn{1}{c}{ $j=2$} & \multicolumn{1}{c}{    $5.290 \times 10^{-11}$  } & \multicolumn{1}{c}{  0.8122  } & \multicolumn{1}{r}{  $-$3.4   }\\\cline{2-5} \cline{8-11}
& \multicolumn{1}{c}{ $j=3$} & \multicolumn{1}{c}{    $3.277 \times 10^{-14}$  } & \multicolumn{1}{c}{  3.5936  } & \multicolumn{1}{r}{  296.4  } & & & \multicolumn{1}{c}{ $j=3$} & \multicolumn{1}{c}{    $5.541 \times 10^{-11}$  } & \multicolumn{1}{c}{  0.8329  } & \multicolumn{1}{r}{  10.3   }\\\cline{2-5} \cline{8-11}
& \multicolumn{1}{c}{ $j=4$} & \multicolumn{1}{c}{    $5.233 \times 10^{-14}$  } & \multicolumn{1}{c}{  3.4386  } & \multicolumn{1}{r}{  308.0  } & & & \multicolumn{1}{c}{ $j=4$} & \multicolumn{1}{c}{    $6.032 \times 10^{-11}$  } & \multicolumn{1}{c}{  0.8214  } & \multicolumn{1}{r}{  49.8   }\\\cline{2-5} \cline{8-11}
& \multicolumn{1}{c}{ $j=5$} & \multicolumn{1}{c}{    $6.747 \times 10^{-14}$  } & \multicolumn{1}{c}{  3.3745  } & \multicolumn{1}{r}{  307.5  } & & & \multicolumn{1}{c}{ $j=5$} & \multicolumn{1}{c}{    $6.902 \times 10^{-11}$  } & \multicolumn{1}{c}{  0.7596  } & \multicolumn{1}{r}{  37.8   }\\\cline{2-5} \cline{8-11}
& \multicolumn{1}{c}{ $j=6$} & \multicolumn{1}{c}{    $7.963 \times 10^{-14}$  } & \multicolumn{1}{c}{  3.3506  } & \multicolumn{1}{r}{  303.9  } & & & \multicolumn{1}{c}{ $j=6$} & \multicolumn{1}{c}{    $8.384 \times 10^{-11}$  } & \multicolumn{1}{c}{  0.6871  } & \multicolumn{1}{r}{  62.7   }\\\cline{2-5} \cline{8-11}
& \multicolumn{1}{c}{ $j=7$} & \multicolumn{1}{c}{    $2.245 \times 10^{-13}$  } & \multicolumn{1}{c}{  2.9066  } & \multicolumn{1}{r}{  554.3  } & & & \multicolumn{1}{c}{ $j=7$} & \multicolumn{1}{c}{    $1.083 \times 10^{-10}$  } & \multicolumn{1}{c}{  0.5929  } & \multicolumn{1}{r}{  69.8   }\\\cline{2-5} \cline{8-11}
& \multicolumn{1}{c}{ $j=8$} & \multicolumn{1}{c}{    $3.009 \times 10^{-13}$  } & \multicolumn{1}{c}{  2.8003  } & \multicolumn{1}{r}{  535.9  } & & & \multicolumn{1}{c}{ $j=8$} & \multicolumn{1}{c}{    $1.237 \times 10^{-10}$  } & \multicolumn{1}{c}{  0.5479  } & \multicolumn{1}{r}{  10.4   }\\\cline{2-5} \cline{8-11}
& \multicolumn{1}{c}{ $j=9$} & \multicolumn{1}{c}{    $3.708 \times 10^{-13}$  } & \multicolumn{1}{c}{  2.7379  } & \multicolumn{1}{r}{  474.3  } & & & \multicolumn{1}{c}{ $j=9$} & \multicolumn{1}{c}{    $1.410 \times 10^{-10}$  } & \multicolumn{1}{c}{  0.5093  } & \multicolumn{1}{r}{  $-$0.5   }\\\cline{2-5} \cline{8-11}
& \multicolumn{1}{c}{ $j=10$} & \multicolumn{1}{c}{    $4.796 \times 10^{-13}$ } & \multicolumn{1}{c}{   2.6591 } & \multicolumn{1}{r}{   404.5 } & & & \multicolumn{1}{c}{ $j=10$} & \multicolumn{1}{c}{    $1.547 \times 10^{-10}$ } & \multicolumn{1}{c}{   0.4875 } & \multicolumn{1}{r}{   2.9   }\\\cline{2-5} \cline{8-11}
\hline
\multirow{11}{*}{$v=2$} & \multicolumn{1}{c}{ $j=0$} & \multicolumn{1}{c}{    $2.308 \times 10^{-13}$  } & \multicolumn{1}{c}{  2.7935  } & \multicolumn{1}{r}{  319.1  } & & \multirow{11}{*}{$v=6$} & \multicolumn{1}{c}{ $j=0$} & \multicolumn{1}{c}{    $6.124 \times 10^{-11}$  } & \multicolumn{1}{c}{  0.8417  } & \multicolumn{1}{r}{  -9.7   }\\\cline{2-5} \cline{8-11}
& \multicolumn{1}{c}{ $j=1$} & \multicolumn{1}{c}{    $6.629 \times 10^{-13}$  } & \multicolumn{1}{c}{  2.3507  } & \multicolumn{1}{r}{  475.4  } & & & \multicolumn{1}{c}{ $j=1$} & \multicolumn{1}{c}{    $8.904 \times 10^{-11}$  } & \multicolumn{1}{c}{  0.6775  } & \multicolumn{1}{r}{  $-$24.5  }\\\cline{2-5} \cline{8-11}
& \multicolumn{1}{c}{ $j=2$} & \multicolumn{1}{c}{    $1.993 \times 10^{-12}$  } & \multicolumn{1}{c}{  1.8977  } & \multicolumn{1}{r}{  610.9  } & & & \multicolumn{1}{c}{ $j=2$} & \multicolumn{1}{c}{    $1.220 \times 10^{-10}$  } & \multicolumn{1}{c}{  0.5626  } & \multicolumn{1}{r}{  1.1    }\\\cline{2-5} \cline{8-11}
& \multicolumn{1}{c}{ $j=3$} & \multicolumn{1}{c}{    $3.118 \times 10^{-12}$  } & \multicolumn{1}{c}{  1.7319  } & \multicolumn{1}{r}{  600.6  } & & & \multicolumn{1}{c}{ $j=3$} & \multicolumn{1}{c}{    $1.243 \times 10^{-10}$  } & \multicolumn{1}{c}{  0.5739  } & \multicolumn{1}{r}{  $-$6.8   }\\\cline{2-5} \cline{8-11}
& \multicolumn{1}{c}{ $j=4$} & \multicolumn{1}{c}{    $2.491 \times 10^{-12}$  } & \multicolumn{1}{c}{  1.8734  } & \multicolumn{1}{r}{  482.1  } & & & \multicolumn{1}{c}{ $j=4$} & \multicolumn{1}{c}{    $1.225 \times 10^{-10}$  } & \multicolumn{1}{c}{  0.5886  } & \multicolumn{1}{r}{  $-$30.5  }\\\cline{2-5} \cline{8-11}
& \multicolumn{1}{c}{ $j=5$} & \multicolumn{1}{c}{    $1.890 \times 10^{-12}$  } & \multicolumn{1}{c}{  2.0457  } & \multicolumn{1}{r}{  397.2  } & & & \multicolumn{1}{c}{ $j=5$} & \multicolumn{1}{c}{    $1.365 \times 10^{-10}$  } & \multicolumn{1}{c}{  0.5445  } & \multicolumn{1}{r}{  $-$35.3  }\\\cline{2-5} \cline{8-11}
& \multicolumn{1}{c}{ $j=6$} & \multicolumn{1}{c}{    $1.905 \times 10^{-12}$  } & \multicolumn{1}{c}{  2.0860  } & \multicolumn{1}{r}{  382.1  } & & & \multicolumn{1}{c}{ $j=6$} & \multicolumn{1}{c}{    $1.487 \times 10^{-10}$  } & \multicolumn{1}{c}{  0.5239  } & \multicolumn{1}{r}{  $-$35.0  }\\\cline{2-5} \cline{8-11}
& \multicolumn{1}{c}{ $j=7$} & \multicolumn{1}{c}{    $3.816 \times 10^{-12}$  } & \multicolumn{1}{c}{  1.7968  } & \multicolumn{1}{r}{  533.1  } & & & \multicolumn{1}{c}{ $j=7$} & \multicolumn{1}{c}{    $1.729 \times 10^{-10}$  } & \multicolumn{1}{c}{  0.4752  } & \multicolumn{1}{r}{  $-$23.5  }\\\cline{2-5} \cline{8-11}
& \multicolumn{1}{c}{ $j=8$} & \multicolumn{1}{c}{    $6.600 \times 10^{-12}$  } & \multicolumn{1}{c}{  1.5654  } & \multicolumn{1}{r}{  591.6  } & & & \multicolumn{1}{c}{ $j=8$} & \multicolumn{1}{c}{    $2.050 \times 10^{-10}$  } & \multicolumn{1}{c}{  0.4161  } & \multicolumn{1}{r}{  $-$29.7  }\\\cline{2-5} \cline{8-11}
& \multicolumn{1}{c}{ $j=9$} & \multicolumn{1}{c}{    $8.468 \times 10^{-12}$  } & \multicolumn{1}{c}{  1.4712  } & \multicolumn{1}{r}{  547.7  } & & & \multicolumn{1}{c}{ $j=9$} & \multicolumn{1}{c}{    $2.280 \times 10^{-10}$  } & \multicolumn{1}{c}{  0.3848  } & \multicolumn{1}{r}{  $-$33.5  }\\\cline{2-5} \cline{8-11}
& \multicolumn{1}{c}{ $j=10$} & \multicolumn{1}{c}{    $9.667 \times 10^{-12}$ } & \multicolumn{1}{c}{   1.4332 } & \multicolumn{1}{r}{   463.3 } & & & \multicolumn{1}{c}{ $j=10$} & \multicolumn{1}{c}{    $2.422 \times 10^{-10}$ } & \multicolumn{1}{c}{   0.3759 } & \multicolumn{1}{r}{   $-$27.0 }\\\cline{2-5} \cline{8-11}
\hline
\multirow{11}{*}{$v=3$} & \multicolumn{1}{c}{ $j=0$} & \multicolumn{1}{c}{    $2.949 \times 10^{-12}$  } & \multicolumn{1}{c}{  1.8080  } & \multicolumn{1}{r}{  279.0  } & & \multirow{11}{*}{$v=7$}& \multicolumn{1}{c}{ $j=0$} & \multicolumn{1}{c}{    $1.686 \times 10^{-10}$  } & \multicolumn{1}{c}{  0.5093  } & \multicolumn{1}{r}{  -5.7   }\\\cline{2-5} \cline{8-11}
& \multicolumn{1}{c}{ $j=1$} & \multicolumn{1}{c}{    $8.269 \times 10^{-12}$  } & \multicolumn{1}{c}{  1.3639  } & \multicolumn{1}{r}{  417.9  } & & & \multicolumn{1}{c}{ $j=1$} & \multicolumn{1}{c}{    $1.829 \times 10^{-10}$  } & \multicolumn{1}{c}{  0.4753  } & \multicolumn{1}{r}{  $-$28.7  }\\\cline{2-5} \cline{8-11}
& \multicolumn{1}{c}{ $j=2$} & \multicolumn{1}{c}{    $1.634 \times 10^{-11}$  } & \multicolumn{1}{c}{  1.0928  } & \multicolumn{1}{r}{  453.1  } & & & \multicolumn{1}{c}{ $j=2$} & \multicolumn{1}{c}{    $1.971 \times 10^{-10}$  } & \multicolumn{1}{c}{  0.4509  } & \multicolumn{1}{r}{  $-$27.7  }\\\cline{2-5} \cline{8-11}
& \multicolumn{1}{c}{ $j=3$} & \multicolumn{1}{c}{    $1.633 \times 10^{-11}$  } & \multicolumn{1}{c}{  1.1446  } & \multicolumn{1}{r}{  403.6  } & & & \multicolumn{1}{c}{ $j=3$} & \multicolumn{1}{c}{    $2.007 \times 10^{-10}$  } & \multicolumn{1}{c}{  0.4486  } & \multicolumn{1}{r}{  $-$32.0  }\\\cline{2-5} \cline{8-11}
& \multicolumn{1}{c}{ $j=4$} & \multicolumn{1}{c}{    $1.482 \times 10^{-11}$  } & \multicolumn{1}{c}{  1.2310  } & \multicolumn{1}{r}{  407.4  } & & & \multicolumn{1}{c}{ $j=4$} & \multicolumn{1}{c}{    $2.088 \times 10^{-10}$  } & \multicolumn{1}{c}{  0.4367  } & \multicolumn{1}{r}{  $-$37.6  }\\\cline{2-5} \cline{8-11}
& \multicolumn{1}{c}{ $j=5$} & \multicolumn{1}{c}{    $1.296 \times 10^{-11}$  } & \multicolumn{1}{c}{  1.3220  } & \multicolumn{1}{r}{  373.7  } & & & \multicolumn{1}{c}{ $j=5$} & \multicolumn{1}{c}{    $2.292 \times 10^{-10}$  } & \multicolumn{1}{c}{  0.4004  } & \multicolumn{1}{r}{  $-$29.9  }\\\cline{2-5} \cline{8-11}
& \multicolumn{1}{c}{ $j=6$} & \multicolumn{1}{c}{    $1.582 \times 10^{-11}$  } & \multicolumn{1}{c}{  1.2313  } & \multicolumn{1}{r}{  344.8  } & & & \multicolumn{1}{c}{ $j=6$} & \multicolumn{1}{c}{    $2.390 \times 10^{-10}$  } & \multicolumn{1}{c}{  0.3966  } & \multicolumn{1}{r}{  $-$25.6  }\\\cline{2-5} \cline{8-11}
& \multicolumn{1}{c}{ $j=7$} & \multicolumn{1}{c}{    $3.020 \times 10^{-11}$  } & \multicolumn{1}{c}{  0.9518  } & \multicolumn{1}{r}{  442.9  } & & & \multicolumn{1}{c}{ $j=7$} & \multicolumn{1}{c}{    $2.644 \times 10^{-10}$  } & \multicolumn{1}{c}{  0.3703  } & \multicolumn{1}{r}{  $-$11.1  }\\\cline{2-5} \cline{8-11}
& \multicolumn{1}{c}{ $j=8$} & \multicolumn{1}{c}{    $3.933 \times 10^{-11}$  } & \multicolumn{1}{c}{  0.8634  } & \multicolumn{1}{r}{  443.7  } & & & \multicolumn{1}{c}{ $j=8$} & \multicolumn{1}{c}{    $3.021 \times 10^{-10}$  } & \multicolumn{1}{c}{  0.3258  } & \multicolumn{1}{r}{  $-$2.9   }\\\cline{2-5} \cline{8-11}
& \multicolumn{1}{c}{ $j=9$} & \multicolumn{1}{c}{    $4.666 \times 10^{-11}$  } & \multicolumn{1}{c}{  0.8169  } & \multicolumn{1}{r}{  426.5  } & & & \multicolumn{1}{c}{ $j=9$} & \multicolumn{1}{c}{    $3.261 \times 10^{-10}$  } & \multicolumn{1}{c}{  0.3056  } & \multicolumn{1}{r}{  2.5    }\\\cline{2-5} \cline{8-11}
& \multicolumn{1}{c}{ $j=10$} & \multicolumn{1}{c}{    $4.734 \times 10^{-11}$ } & \multicolumn{1}{c}{   0.8362 } & \multicolumn{1}{r}{   360.0 } & & & \multicolumn{1}{c}{ $j=10$} & \multicolumn{1}{c}{    $3.319 \times 10^{-10}$ } & \multicolumn{1}{c}{   0.3127 } & \multicolumn{1}{r}{   9.5   }\\\cline{2-5} \cline{8-11}
\hline
\end{tabular}
\tablenotea{\\The parameters $\alpha$, $\beta$, and $\gamma$ are obtained by fitting the QM/QCT rate coefficients (provided as Supplementary Information) to the expression $k(T) = \alpha (T/300)^{\beta} \exp (- \gamma / T )$, where $T$ is the temperature in degrees Kelvin. The fits are valid for the temperature range 100-3000\,K, except for $v$ = 0, $j$ = 0-3, in which case the fits are only valid in the 200-3000\,K temperature range.}
\end{table*}

To make realistic abundance predictions of the product XH, PDR models need to determine the H$_2$\,($v$,\,$j$) level populations (typically very far from thermal) and include state-specific reaction rates in their chemical networks \mbox{\citep[][]{Agundez10}}. These state-dependent reaction rates can either be measured in the laboratory \citep[e.g.,][]{Hierl_1997} or determined through \mbox{\textit{ab initio}} calculations of potential energy surfaces (PES), followed by a study of the scattering process. Recent calculations of state-specific rates for hydrogen abstraction reactions with \mbox{X\,=\,C$^+$} \citep{Zanchet13b}, \mbox{X\,=\,S$^+$} \citep{Zanchet13a,Zanchet19}, and \mbox{X\,=\,S and SH$^+$} \citep{Goicoechea21} allow to explain the abundances of CH$^+$, SH$^+$, and SH inferred from observations of the Orion Bar PDR \citep[e.g.,][]{Nagy13,Joblin18, Goicoechea21}.

However, the reactivity of \mbox{H$_2$\,($v,j$)\,+\,X\,$\rightarrow$\,XH\,+\,X} collisions is a very selective process, often hard to estimate by simple educated guesses \citep{jambrinapccp12}. In this study we carry out \textit{ab initio} quantum calculations of reaction~(\ref{reac-1}), which is endothermic by $E/k \sim 770$\, K \citep{Huber1979,Baulch2005,Ruscic01,Joens01} and has an even higher vibrationally adiabatic barrier from H$_2$\,($v=0,j=0$) of $V_0^\ddagger$ = 5700\,K ($E_b/k$ = 6900\,K without the respective zero point energies; \citealt{zanchetjcp19}.) After determining the H$_2$ state-dependent reaction rate coefficients, we investigate whether reaction of O\,($^3P$) with \mbox{FUV-pumped} vibrationally excited H$_2$ enhances the formation of OH in strongly irradiated PDRs, compared to models that simply use the thermal reaction rate coefficient. We do this by implementing PDR models adapted to the physical and illumination conditions of the Orion Bar, where the \textit{Herschel} space telescope detected several far-infrared OH rotational lines \citep{Goicoechea11}.

In Sec.~\ref{sec:quantum} we give details of the PES and scattering calculations and how we obtained the state-specific rate coefficients of reaction~(\ref{reac-1}), while in Sec.~\ref{sec:model} we investigate the impact of the calculated rate coefficients on PDR models adapted to the Orion Bar.

\section{Quantum calculations} \label{sec:quantum}

Three PESs, two of symmetries $^3A''$ and one of symmetry $^3A'$ correlate adiabatically with the ground state of the reactants. Two of them, $1^3A'$ and $1^3A''$, also correlate with the ground electronic state of the products and are necessary to describe the \mbox{O\,($^3P$)\,+\,H$_2$} reaction. The two PESs are degenerate for collinear geometries, where the saddle point is found, at $E_b/k$ = 6858\,K with respect to the reactants asymptote \citep{zanchetjcp19}, while for non-collinear approaches the barrier is consistently lower on the $^3A''$ PES. We have performed adiabatic scattering calculations on both PESs for \mbox{O\,($^3P$)\,+\,H$_2$}\,($v \leq$7, $j \leq$10) to determine state-selected reaction rates between 100\,K and 3500\,K using the procedure described below. Time-independent quantum mechanical (QM) scattering calculations were carried out applying the coupled-channel hyperspherical method implemented in the ABC code \citep{SCM:CPC00} to the set of PESs calculated by \cite{zanchetjcp19}. Each PES consists of 5000 \textit{ab initio} energies calculated using the internally contracted multireference configuration interaction method (icMRCI) including simple and double excitation, Davidson correction, and a aug-c-pV5Z basis set. The fits to the PESs obtained account for the degeneracy of the collinear saddle point very accurately \citep{zanchetjcp19}. They have been used by \cite{JZABA:NC16} to simulate the experimental $\Lambda$-doublet populations of nascent OD with an excellent agreement with the experimental results of \cite{LZMM:NC13,LZMM:JACS14}. It was also possible to explain the propensity towards $\Pi(A')$ \mbox{$\Lambda$-doublet} states as a consequence of the presence of a mechanism on the $A''$ PES that entails a change in the direction of the doubly occupied orbital of atomic oxygen \citep{JZABA:NC16,JambrinaPCCP19}.

QM scattering calculations on the $1^3A'$ and $1^3A''$ PESs were carried out at 60 total energies between 0.3 eV and 2.5 eV including all partial waves ($J$) to convergence ($J_{\rm max}=62$ at the highest energies) and all helicity projections up to 26. Propagation was carried out in 300 log-derivative steps up to a hyperradius of 20 $a_0$ including in the basis all the diatomic energy levels up to 3.25 eV. For $v=4, 5$ $j=0$ states additional QM calculations were run at total energies slightly above the opening of vibrational states $v = 4,5$ of H$_2$. The excitation functions $\sigma_{_R}(E_{\rm coll})$ for the \mbox{O\,($^3P$)\,+\,H$_2$\,($v,j$)} calculated on both PESs exhibit significant thresholds up to $v=4$. As could be expected, the higher the internal energy of H$_2$, the smaller becomes the energy threshold. However, only a fraction of the internal energy of H$_2$ is used to surmount the barrier, and even though the energy of the ($v=1,j=0$) state of H$_2$ is already above the electronic barrier, the reaction only becomes barrierless for H$_2$\,($v \geq 4$). Since the reaction is barrierless for $v \geq 4$, state-specific reaction rates for H$_2$\,($v \geq 4$) are expected to be significantly large even at low temperatures.

\begin{figure}
\centering
\includegraphics[angle=0,width=\columnwidth]{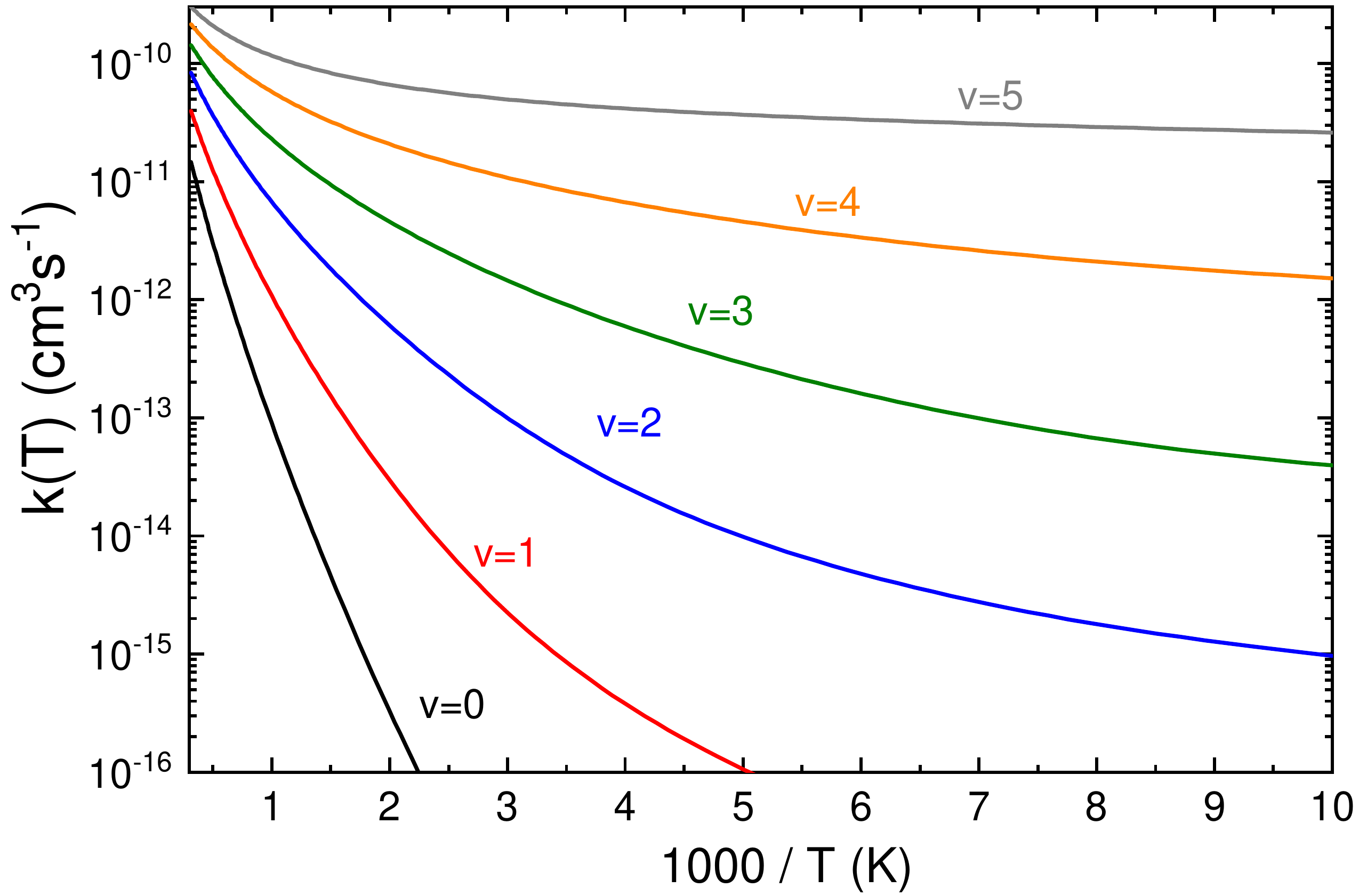}
\caption{Calculated rate coefficients of the reaction O\,($^3P$) + H$_2$\,($v,j=0$) for selected vibrational states $v$ of H$_2$.} \label{fig1}
\end{figure}

To calculate the state-specific reaction rate coefficients up to temperatures of 3500 K, it was necessary to calculate integral cross sections for collision energies up to 2.5-3.0 eV. Since converged QM calculations are computationally very expensive for hot rovibrational states of H$_2$ at these very high energies, we complemented the QM scattering calculations with batches of $2 \times 10^6$ quasiclassical trajectories (QCT) per rovibrational state that were calculated using the procedure of \cite{AHR:JCP92,ABH:JCSFT98}. In these batches of trajectories, the initial atom-diatom distance was set at 20 \AA, and an integration step of 0.02 fs was used, guaranteeing a total energy conservation better than one part in $10^5$. QCT calculations systematically predicted a larger energy threshold, but at the energies in which QCT calculations were merged with the QM ones, the agreement between the two sets of calculations was good, so any effect in the rate coefficients consequence of this merge should be negligible. The very high internal energy of H$_2$\,($v= 6,7$) impair to carry out QM scattering calculations for these states, and thus we relied on QCT calculations. Since for H$_2$\,($v= 4, 5$) the reaction is already barrierless, rate coefficients predicted by the QCT method for H$_2$\,($v= 6,7$) should be very similar to those obtained through QM calculations.

Once the state-specific rate coefficients were determined on the $^3A'$ and $^3A''$ PESs, the overall state-specific rate coefficients were calculated as follows:
\begin{equation} \label{rate}
k(T) = \frac{3  k_{A''}(T) + \left[ 2 + \exp\big(-\frac{\Delta E_1}{T}\big)
\right] k_{A'}(T)}{5 + 3\cdot \exp\big(-\frac{\Delta E_1}{T}\big) +
\exp\big(-\frac{\Delta E_0}{T}\big)},
\end{equation}
where $T$ is the temperature, $k(T)$ is the overall state-specific rate coefficient, $k_{A'}(T)$ and $k_{A''}(T)$ are the state-specific rate coefficients on the $A'$ and $A''$ PESs, respectively. Equation~(\ref{rate}) accounts for the O\,($^3P$) spin-orbit splitting, where $\Delta E_1$=227.708 K and $\Delta E_0$=326.569 K are the energies of the spin-orbit states $^3P_1$ and $^3P_0$, respectively, over the ground state $^3P_2$. Equation~(\ref{rate}) also considers the correlation between the three levels of O\,($^3P$) and the $A'$ and $A''$ PESs. As established by \textit{ab initio} calculations including spin-orbit correction along the reaction path, the 1$^3A''$ state correlates with three of the five components of O\,($^3P_2$), and the 1$^3A'$ PES correlates with the other two components of O\,($^3P_2$) and one component of O\,($^3P_{1}$). The remaining three components of O\,($^3P_{1,0}$) correlate with the non-reactive 2$^3A''$ state.

In Fig.~\ref{fig1} we show the state-specific rate coefficients for the reaction \mbox{O\,($^3P$)\,+\,H$_2$\,($v=0$-$5,j=0$)}. It is clear from the figure that vibrational excitation of H$_2$ leads to a significant increase of $k(T)$, especially at low temperatures. For example, at 500 K the ratio between the rate coefficients for $v=4$ and $v=0$ is $7.5 \times 10^4$, while at 1000 K this ratio is just 640. Interestingly, even for $v=5$ the rate coefficient is not independent of temperature, as it would be expected for a barrierless reaction. For $v$ = 0-1, our results are in good agreement with those calculated by \cite{Bala04} on the PES of \cite{Rogers00}, which predicted a somewhat smaller collinear barrier. Also in agreement with previous results \citep{Weck06}, we also predict the existence of an energy barrier for $v=2$-3. However, to the best of our knowledge there are not previous calculations for $v > 3$ for which the reaction proceeds without no barrier.

\section{Impact on the abundance of OH} \label{sec:model}

\begin{figure}
\centering
\includegraphics[angle=0,width=\columnwidth]{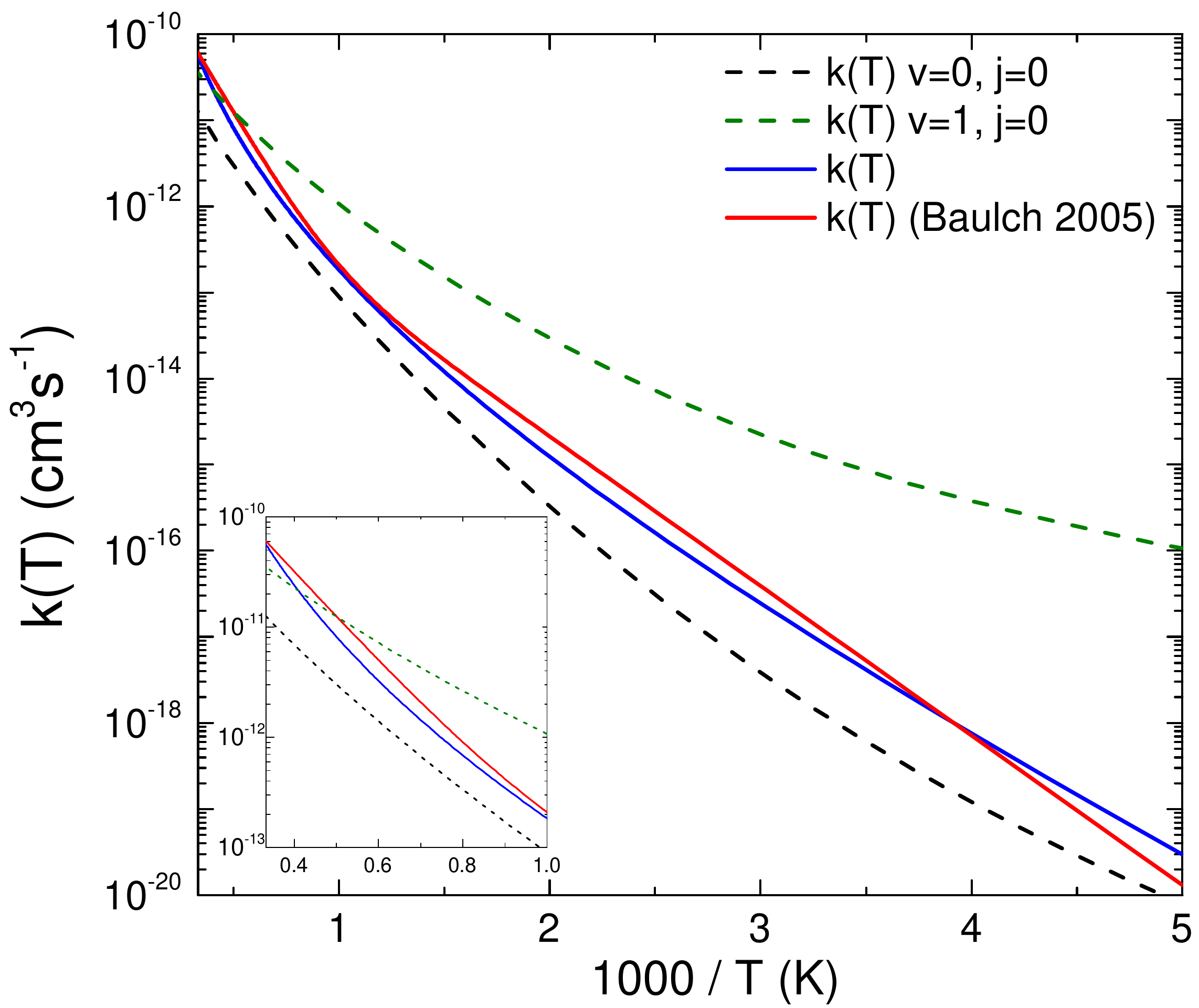}
\caption{Thermal rate coefficient calculated in this work, given by Eq.~(\ref{rate_thermal}), is compared with the experimental one \citep{Baulch2005}. Rate coefficients for H$_2$ in the ($v=0,j=0$) and ($v=1,j=0$) states are also shown for the sake of comparison. The inset highlights the behavior at high temperatures ($>$\,1000 K).} \label{fig2}
\end{figure}

\begin{figure*}
\centering
\includegraphics[angle=0,width=0.95\textwidth]{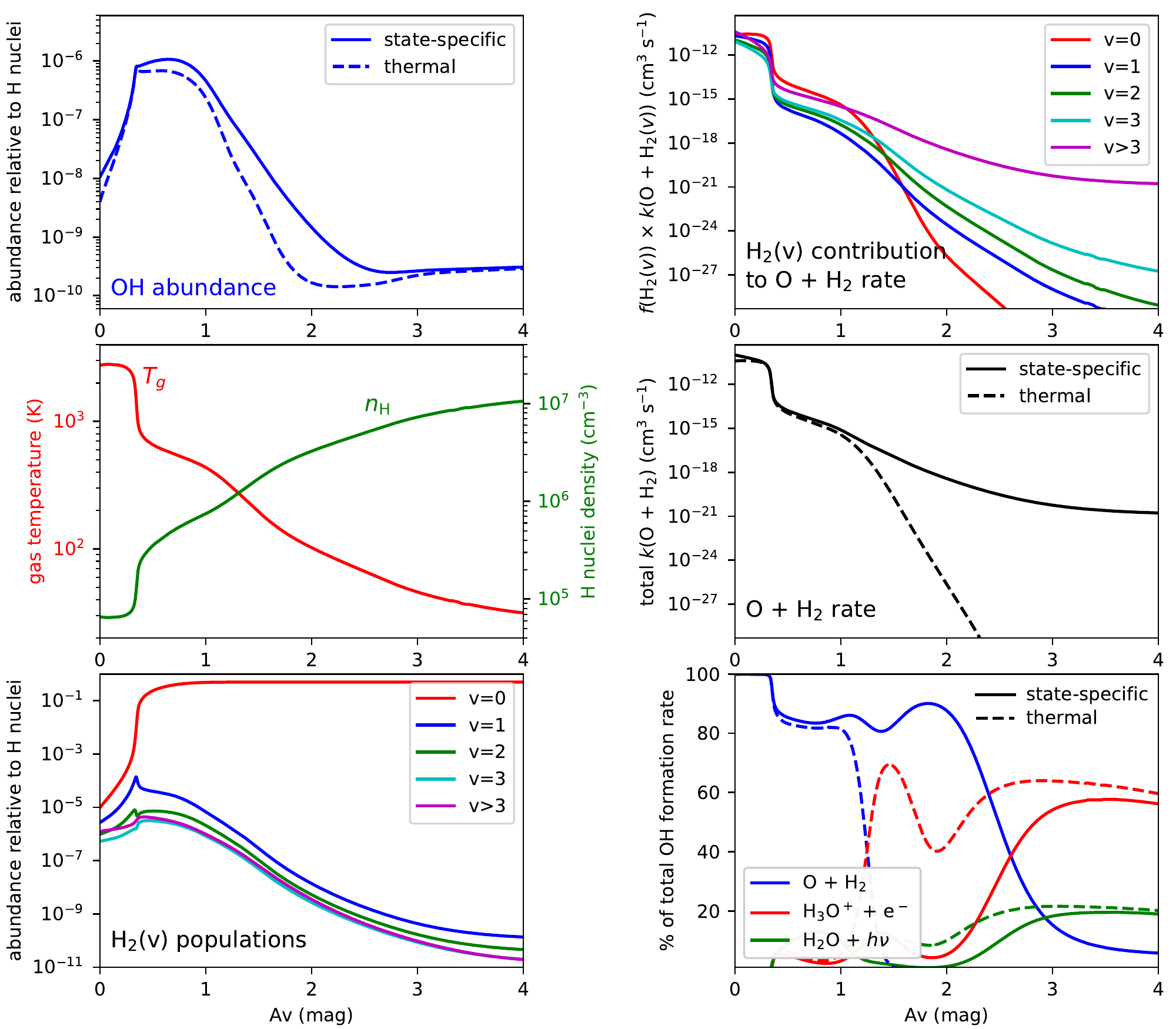}
\caption{Assorted quantities calculated in the PDR model of the Orion Bar plotted as a function of $A_{\rm V}$. Solid lines refer to the model including H$_2$ state-specific rate coefficients for reaction~(\ref{reac-1}), while dashed lines correspond to the reference model, in which a thermal rate coefficient is used for reaction~(\ref{reac-1}). The top right panel shows the contribution of each vibrational state of H$_2$ to the total rate coefficient of the O + H$_2$ reaction, expressed as $f$(H$_2$($v$)) $\times$ $k$(O + H$_2$($v$)), where $f$(H$_2$($v$)) is the fractional population of H$_2$ in the vibrational state $v$. The bottom right panel shows the contribution of the three main gas-phase reactions of formation of OH, expressed as percentage of the total OH formation rate.}
\label{fig:model}
\end{figure*}

Previous studies of the impact of reactions of vibrationally excited H$_2$ on the chemistry of molecular clouds have shown that the most dramatic effect of
including state-dependent rate coefficients is attained in dense and strongly UV-illuminated PDRs like the Orion Bar
\citep{Agundez10,Zanchet13a,Zanchet13b,Zanchet19,Goicoechea21}. To evaluate the role of FUV-pumped vibrationally excited H$_2$ reacting with atomic oxygen on the abundance of OH we have therefore focused on the Orion Bar PDR.

We have used the Meudon PDR code\footnote{\texttt{https://ism.obspm.fr/}} \citep{LePetit06}  in which we have implemented the \mbox{H$_2$\,($v,j$)} state-specific rate coefficient expressions for reaction~(\ref{reac-1}) given in Table~\ref{table:rates}. We assume a cloud having an isobaric structure with a thermal pressure \mbox{($P_{th}$ = $n_{\rm H}\,T$)} of $2 \times 10^8$ cm$^{-3}$ K, a total visual extinction of 10 mag, a FUV radiation field 10$^4$ times stronger than the interstellar radiation field of \cite{Draine78} illuminating on one side of the cloud, and a cosmic-ray ionization rate of H$_2$ of 10$^{-16}$ s$^{-1}$. The models include only gas-phase chemistry and formation of H$_2$ on grain surfaces \citep{LeBourlot2012}. To rigorously evaluate how is the OH abundance affected by the FUV pumped non-thermal H$_2$ population distribution in the reaction with O we have run a reference model in which we use a thermal rate coefficient for reaction~(\ref{reac-1}), given by
\begin{equation} \label{rate_thermal}
k(T) = 2.22 \times 10^{-14} {\rm cm^3 s^{-1}} (T/300)^{3.75} \exp{(-2401/T)},
\end{equation}
where the temperature $T$ is in units of degrees Kelvin. Equation~(\ref{rate_thermal}) is obtained from a fit to the thermal rate coefficient calculated from the individual rate coefficient expressions for each ($v,j$) state of H$_2$ given in Table~\ref{table:rates}. Therefore, any difference in the OH abundance between the model including state-specific rates and the reference model can be confidently ascribed to the effect of the non-thermal population distribution of H$_2$ on the reactivity with O. In Fig.~\ref{fig2}, we compare the thermal rate coefficient expression given by Eq.~(\ref{rate_thermal}) with the experimental expression given in the compilation of \cite{Baulch2005} and, as it is apparent from the figure, both are in a very good agreement. In Fig.~\ref{fig2} we also compare the thermal rate coefficient with the state-specific rate coefficients calculated for the ($v=0,j=0$) and ($v=1,j=0$) states of H$_2$. Even at low temperatures, excited rotational levels of H$_2$ have non-negligible thermal populations and the increase in the rate coefficient with the quantum number $j$ of H$_2$ is enough to make the thermal rate coefficient higher than its ($v=0,j=0$) counterpart. In any case, at low temperatures the thermal rate coefficient is still well below the rate coefficient of the ($v=1,j=0$) state of H$_2$.

In the top left panel of Fig.~\ref{fig:model} we show the calculated abundance of OH in the edge ($A_{\rm V}$ = 0-4) of the Orion Bar. It is seen that OH reaches a maximum fractional abundance relative to H nuclei of $\sim 10^{-6}$ in the $A_{\rm V}$ = 0.5-1.0 range, decreasing at lower $A_{\rm V}$ due to the high FUV flux, which dissociates molecules, and also at high $A_{\rm V}$ due to the lower temperatures encountered. The main effect of including state-specific rate coefficients for \mbox{O + H$_2$} occurs in the range $A_{\rm V}$ = 1.3-2.3, where the OH abundance increases by up to one order of magnitude with respect to the reference model. However, this is not the region of maximum abundance of OH and the impact on the column density of OH at $A_{\rm V}$ = 4 is modest, just $\sim60$~\%. The small change in the OH abundance at \mbox{$A_{\rm V} < 1.3$} is related to the fact that in this region temperatures are high (see middle left panel in Fig.~\ref{fig:model}) and the reaction \mbox{O + H$_2$} occurs fast either if we consider the thermal rate coefficient or if we use state-specific rates. In the reference model, at $A_{\rm V} >1.3$ temperatures decrease and the thermal rate coefficient of reaction \mbox{O + H$_2$} takes small values. In these more internal regions, the reaction between O and H$_2$ is no longer the main route to OH. Our pure gas-phase models point to the ion H$_3$O$^+$ as the main precursor of OH (through dissociative recombination with electrons) in these internal regions (see bottom right panel in Fig.~\ref{fig:model}), although this pathway has a limited efficiency resulting in a relatively low abundance of OH. However, when \mbox{O + H$_2$} state-specific rates are considered and the non-thermal population distribution of H$_2$ is taken into account, reaction~(\ref{reac-1}) becomes the main route to OH even in deeper regions, up to $A_{\rm V}$ = 2.6, enhancing the abundance of OH. FUV pumping makes H$_2$ to maintain a certain population of $v \geq 1$ levels even at relatively deep regions of the cloud. As shown in the middle panel of Fig.~\ref{fig:model}, the fractional populations of $v \geq 1$ levels are around 10$^{-7}$ at $A_{\rm V}$ = 1.3 and they decrease smoothly with increasing $A_{\rm V}$. The highest contribution to the formation of OH is provided by H$_2$ $v > 3$ levels (see second panel from bottom in Fig.~\ref{fig:model}), for which the \mbox{O + H$_2$} reaction starts to attain high rate coefficients, close to the collision limit (see Fig.~\ref{fig1}). The increase in the abundance of OH at $A_{\rm V} > 1.3$ when implementing state-specific rates is clearly a consequence of an enhancement in the \mbox{O + H$_2$} reaction rate. As shown in the bottom panel of Fig.~\ref{fig:model}, the \mbox{O + H$_2$} thermal rate coefficient vanishes down to very small values at increasing $A_{\rm V}$ following the decrease in temperature, while non-thermal population of vibrational levels of H$_2$ allows to increase the reaction rate by orders of magnitude.

We expect that, when implemented in more refined gas-grain models of PDRs and illuminated protoplanetary disks \citep[\mbox{i.e., including} OH- and H$_2$O-ice mantle formation and desorption, see e.g.,][]{Hollenbach09}, these state-specific rate coefficients will provide more accurate OH abundances and line intensity estimations. Moreover, from our QM and QCT calculations it is possible to extract state-to-state rate coefficients that will allow to include formation pumping in OH excitation models \citep[e.g.,][]{Tabone21} and to accurately model near future observations of OH in these environments with the \textit{James Webb Space Telescope} (JWST).

\begin{acknowledgements}

We acknowledge the anonymous referee for a constructive report that allowed to improve this article. We acknowledge the Spanish Ministerio de Ciencia e Innovaci\'on for funding support through the projects \mbox{AYA2016-75066-C2-1-P}, \mbox{FIS2017-83473-C2}, \mbox{PGC2018-096444-B-I00}, \mbox{PID2019-106110GB-I00}, and \mbox{PID2019-107115GB-C21}. A.V. and P.G.J. acknowledge funding by Fundaci\'on Salamanca City of Culture and Knowledge (programme for attracting scientific talent to Salamanca). M.A. also acknowledges funding support from the Ram\'on y Cajal programme of Spanish Ministerio de Ciencia e Innovaci\'on (grant RyC-2014-16277).

\end{acknowledgements}


\bibliographystyle{aa}
\bibliography{references}

\begin{appendix}\label{Sect:Appendix}

%----------------------------------------------------------------------

\end{appendix}

\end{document}